\documentstyle[preprint,aps,tighten,prl]{revtex}
\begin{document}
\draft
\preprint{UTPT-94-29, CITA-94-47}
\title{Remarks about Static Back-Reaction on Black Hole Spacetimes}
\author{ Neil J. Cornish}
\address{Department of Physics, University of Toronto \\ Toronto,
ON M5S 1A7, Canada}
\author{Janna J. Levin}
\address{Canadian Institute for Theoretical Astrophysics\\
60 St. George Street,
Toronto, ON M5S 1A7, Canada}
\maketitle
\begin{abstract}

Recently, it has been claimed that the back reaction
of vacuum polarization on a black hole spacetime naturally
regularizes infinities in the black hole entropy.
We examine the back reaction calculation and
find no such short-distance cut-off,
in contradiction with these
recent claims.
Moreover, the intuitive expectation that the
perturbative calculation breaks down near the event horizon is confirmed.
The new surface gravity diverges and
the metric is degenerate at the stretched horizon.

\end{abstract}
\narrowtext

It has recently been claimed\cite{l} that the back-reaction
on a black hole spacetime
due to vacuum polarization
provides a cut-off needed to regularize the
entropy.
The 1-loop quantum correction to the entropy from a scalar field
near the event horizon diverges, as shown by 't Hooft \cite{thooft}.
The divergence
results since an infinite number of states can contribute at the horzion.
One might conjecture that if a particle could never
actually reach the event horizon, then
only a finite number of states would contribute to the
entropy and the infinites would be regularized.
In Ref. \cite{l}, Lousto claims to have found
such a brick wall by computing the back reaction due to vacuum
polarization on the spacetime.
In the corrected metric, he argues, the acceleration
required to keep a particle at rest vanishes at some
distance outside the stretched horizon
and at that point a particle can remain in stable equilibrium.
Therefore a natural barrier, and hence a short-distance cut-off, arises.
We critically examine this claim and arrive at contrary
conclusions.

We find that the back-reaction-corrected metric has a
new event horizon, stretched by an amount proportional to the
Planck length outside the unperturbed horizon.
Unlike in Ref. \cite{l}, there is no barrier to reaching the horizon.
The acceleration required to keep a particle stationary
never vanishes. In fact, the acceleration diverges at the stretched horizon.
The new surface gravity diverges as well.
Not surprisingly, we also find that our perturbative calculation fails
before the horizon can be reached. In short, this perturbative
back-reaction calculation is unable to shed light on the physics
near the event horizon.

Following the method employed by Lousto in Ref.\cite{l}, we
consider the back reaction due to a conformally invariant
scalar field in the Boulware vacuum~\cite{bl}. The one-loop renormalised
energy momentum tensor in the spacetime of a Schwarzschild black hole
is~\cite{p,fz} (G=c=1, h$\neq 1$)
\begin{equation}
<B|T_{\mu}^{\nu}|B>^{ren}={\alpha M^4\over r^6}\left\{
{\left(2-3{M\over r}\right)^2\over\left(1-2{M\over r}\right)^2}
(-\delta_{\mu}^{\nu}+4\delta_{\mu}^{0}\delta_{0}^{\nu})
+6(3\delta_{\mu}^{0}\delta_{0}^{\nu}
+\delta_{\mu}^{1}\delta_{1}^{\nu})\right\},
\label{1}
\end{equation}
where
\begin{equation}
\alpha={1\over 1440\pi^2}\left({M_{\text{Pl}}\over M}\right)^2~.
\label{7}
\end{equation}
The expression in Eqn.(\ref{1}) differs from that calculated in the
Hartle-Hawking thermal state~\cite{hh} $|H>$ (often misleadingly referred to
as a vacuum state) by a purely thermal contribution from the bath of Hawking
radiation,
\begin{equation}
<H|T_{\mu}^{\nu}|H>^{ren}-<B|T_{\mu}^{\nu}|B>^{ren}
={\sigma T_{loc}^{4} \over 3}
\left(\delta_{\mu}^{\nu}-4\delta_{\mu}^{0}\delta_{0}^{\nu}\right) \; ,
\end{equation}
where $\sigma$ is the Stephan-Boltzman constant,
$T_{loc}=T_{H}(1-2M/r)^{-1/2}$ is the temperature of the black hole
as measured by a static observer at radial
position $r$ and $T_{H}$ is the usual Hawking temperature.
The Boulware vacuum and the Hartle-Hawking thermal state are
both time-reversal invariant and thus suffer no particle production.
Consequently, we
are only considering the back-reaction due to vacuum polarization.

A more realistic calculation of back-reaction effects would require us to use
the Unruh vacuum~\cite{u} which allows for particle production and
fluxes of energy-momentum due to Hawking radiation. However, we shall confine
our current analysis to back-reaction in the Boulware vacuum in order
to critically examine the results in Ref.\cite{l}.

Choosing the Schwarzschild gauge, the metric for a static spherically
symmetric spacetime is
\begin{equation}
ds^2=-a(r)dt^2+b(r)^{-1}dr^2+r^2d\Omega^2~,~~
d\Omega^2=d\vartheta^2+\sin^2\vartheta d\varphi^2~.\label{2}
\end{equation}
Einstein's equations yield\cite{ls}
\begin{equation}
b(r)=1-2{M_{t}\over r}+{1\over r}\int_{\infty}^r\tilde r^2{<B|T_{t}^{t}|B>
^{ren}d\tilde r}~,
\label{3}
\end{equation}
and
\begin{equation}
a(r)=b(r)\exp\left\{\int_{\infty}^r{<B|T_r^r-T_{t}^{t}|B>^{ren}
{\tilde r\over b(\tilde r)}d\tilde r}\right\}~.
\label{4}
\end{equation}
The quantity $M_{t}$ represents the total mass due to the vacuum energy
and the black hole as measured by asymptotic observers. In order to parallel
the treatment in Ref.\cite{l} we shall assume $M_{t}=M$ where $M$ is
the unperturbed black hole mass. Still, it is unclear that this assumption
can be justified.  Perhaps it could be argued that since no matter has been
added or removed, the mass should not change. However, when using the
Hartle-Hawking thermal state, it is certainly incorrect to equate $M_{t}$
and $M$ since radiation has been added to the system\cite{us}.

Inserting Eq.\ (\ref{1}) into Eq.\ (\ref{3}), with $M_t=M$, gives
\begin{equation}
b(r)=1-2{M\over r}+\alpha\left[-{33 \over 4}{M^4\over r^4}
+{9\over8}{M^3\over r^3}
+{15\over16}{M^2\over r^2}-{3\over 16}{M^2\over r(r-2M)}+{3\over8}
{M\over r}\ln{\left(1-2{M\over r}\right)}\right]~,
\label{5}
\end{equation}
where we differ from Lousto's expression for $b(r)$ in Ref\cite{l}
in the third and sixth terms. It is the sign difference in the sixth term
which accounts for the majority of the differences between our work and
that in Ref\cite{l}. We find that $b(r)$ vanishes at $b(r_{h})=0$
where
\begin{equation}
r_{h}=M\left(2+{\sqrt{3\alpha} \over 4}
-{3\alpha \over 32}\ln{\left({3\alpha
\over 64}\right)}+{9 \alpha \over 64}+\dots\right)\; .
\end{equation}
The leading correction to the position of the event horizon,
$M\sqrt{\alpha}$, is proportional to $M_{{\rm Pl}}$, so
the stretched horizon at $r_{h}$ is roughly a Planck length from the
unperturbed horizon.

In the neighbourhood of $r_{h}$ we find $b(r)$ approaches zero linearly,
\begin{equation}
b(r_{h}+x)={x \over M}\left(1+{\sqrt{3\alpha} \over 8}\ln{\left(
{3\alpha \over 64}\right)}
-{\sqrt{3\alpha} \over 8}+ \dots\right)
+{\cal O}\left({x^2 \over M^2}\right)\; .
\end{equation}
The above equation exhibits the important feature that the limit $\alpha
\rightarrow 0$ is {\em non-analytic} in the neighborhood of $r_{h}$ since
\begin{equation}
\lim_{\alpha\rightarrow 0}b(r_{h}+x)={x \over M}+
{\cal O}\left({x^2 \over M^2}\right)\; ,
\end{equation}
while for the uncorrected metric with $\alpha=0$ we have $r_{h}=2M$ and
\begin{equation}
b_{\alpha=0}(2M)={x \over 2M}\; .
\end{equation}
This is a clear sign that our perturbation theory breaks down at $r_{h}$
as our results are no longer analytic (perturbative) in $\alpha$. This can
easily be understood by considering the energy momentum tensor near the
horizon and noticing that it diverges as $r$ approaches $2M$ due to the
redshift factor to the fourth power. This factor of $(1-2M/r)^{-2}$ is
proportional to $1/\alpha$ at the stretched horizon $r_{h}$, and has the
effect of canceling the overall small expansion parameter $\alpha$.
The vacuum energy is no longer a small perturbation near the horizon
and our approximation scheme breaks down.

Continuing our analysis, despite our misgivings, we proceed to calculate
the metric function $a(r)$.
The integral in Eqn. (\ref{4}) is difficult to evaluate in closed form and
we were only able to obtain explicit expressions for $a(r)$
when $\alpha << 1$ and $r >> r_{h}$. The restriction $r>>r_{h}$ occurs
because the integrand is singular at $r_{h}$. With these restrictions
we find
\begin{eqnarray}
a(r)=&&1-2{M\over r}+{\alpha\over 8\left(1-2{M\over r}\right)}
\Bigg[20{M^5\over r^5}-32{M^4\over r^4}-14{M^3\over r^3}
+24{M^2\over r^2}-6{M\over r}\cr\cr
&& -18{M^2\over r^2}\ln{\left(1-2{M\over r}
\right)}+15{M\over r}\ln{\left(1-2{M\over r}\right)}-
3\ln{\left(1-2{M\over r}\right)}\Bigg]+{\cal O}(\alpha^2),
\label{a1}
\end{eqnarray}
In order to evaluate the integral in the neighborhood of $r_{h}$ we employed
the method of steepest descents and found that $a(r)$ near $r=r_{h}$ is
given by
\begin{equation}
a(r_{h}+x)=C\left({x \over M}\right)
^{(1/3-\sqrt{3\alpha}\ln(3\alpha/64)/12+\dots)}\; .
\label{a2}
\end{equation}
The constant $C$ is an uninteresting numerical factor. We see that $a(r)$
does vanish at $r=r_{h}$, showing that $r_{h}$ is indeed the position of
the new event horizon.

To complement the above analytic expressions for $a(r)$ in the
regions $r>> r_{h}$ and $r \approx r_{h}$ we display a numerical evaluation
of $a(r)$ in Fig. 1. for the choice $\alpha=0.1$. This value of $\alpha$
was chosen to facilitate comparison with Lousto's plot of $a(r)$ in Fig. 1.
of Ref.\cite{l} where by contrast $a(r)$ fails to pass through zero
but rather reaches a minimum before asymptoting to $+\infty$ as $r$
approaches $2M$.

Considering the equations of motion for a radially directed test
particle with unit energy at spatial infinity, we find
the coordinate velocity is given by
\begin{equation}
\left({dr\over dt}\right)^2=a b(1-a).
\end{equation}
According to Ref.\cite{l}, Fig. 1., a ``brick wall'' is encountered
just outside of $r=2M$ at the point where $a(r)=1$ and the coordinate
velocity vanishes. An analogous ``brick
wall'' is also said to exist for strings, membranes and scalar fields,
thus providing an ultraviolet cut-off for entropy calculations. From our
analytic expressions (\ref{a1}), (\ref{a2}), and the numerical evaluation
displayed in our Fig.1. we see $a(r)\leq 1$ and no ``brick wall'' is
encountered.

\begin{figure}[h]
\vspace{60mm}
\includegraphics{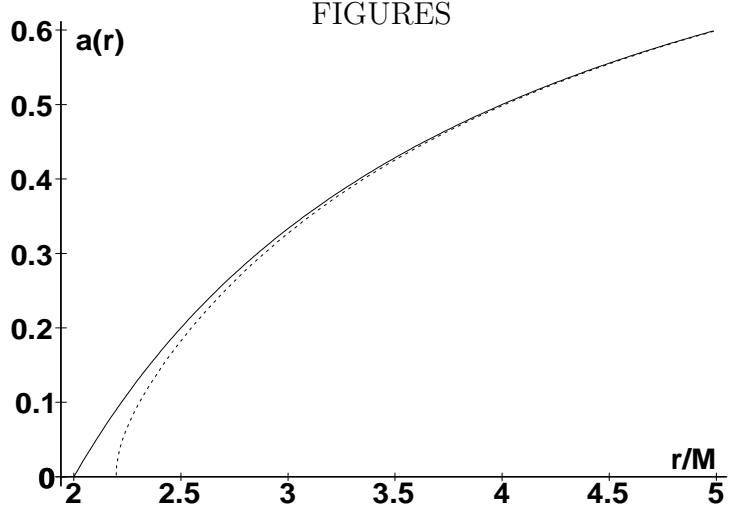}
\vspace{5mm}
\caption{The dashed line shows $a(r)$ in the region between $r=r_{h}$ and
$r=5M$ for the choice $\alpha=0.1$, while the solid line shows $a(r)$ for
the unperturbed Schwarzschild metric}
\end{figure}

It is also stated in Ref.\cite{l} that the coordinate
acceleration, $\ddot{r}=d^2 r/dt^2$, vanishes at the ``brick wall'',
allowing particles to be placed in stable equilibrium ($\dot{r}=
\ddot{r}=0$) just outside the horizon. We disagree with that assertion.
In order to remain stationary against the pull of the black hole,
an observer must be accelerated by amount $A^\mu=\nabla^\mu \ln V$,
where $V$ is the norm of the timelike Killing vector
$\xi_{\mu}=(a(r), 0, 0,0)$.
The magnitude of the acceleration can be written in terms of the metric
components as
\begin{equation}
    A= (A^\mu A_\mu)^{1/2}={a' b^{1/2} \over 2a}\; .
\end{equation}
We see that neither the proper acceleration $A$, nor the coordinate
acceleration $\ddot{r}=aA$ vanish outside of $r=r_{h}$.
Symbolically, using expressions (\ref{3}) and (\ref{4}), the proper
acceleration is
\begin{equation}
A={1\over 2b^{1/2}}
     \left [ b'+    <B| T_r^r-T_t^t |B>\right ]
     \ \ .
\end{equation}
This quantity diverges on the horizon since $b$ approaches zero while
the term in brackets is finite and non-vanishing. Explicitly,
we find $A(r_{h}+x)\propto x^{-1/2}$ and
$\ddot{r}(r_{h}+x) \propto x^{-1/6}$ to leading order in $\alpha$.
The acceleration not only does not vanish, but actually diverges near
$r=2M$.

To complete our analysis we shall calculate the surface gravity
and the metric determinant in our
back-reaction corrected spacetime. The break down of our calculation
at the stretched horizon will be apparent in these quantities.

The surface gravity is the force which would need to be
exerted at infinity in order to keep an observer stationary
at the event horizon:
     \begin{equation}
     \kappa\equiv lim_{r \rightarrow r_h} \left [ V(A^\mu A_\mu)^{1/2}
     \right ]={a'\over 2}\left ({b\over a}\right )^{1/2}|_{r=r_h} \; .
     \end{equation}
The surface gravity is then
     \begin{equation}
     \kappa=
     \left ( b'+<B|T_r^r-T_t^t|B>r\right )|_{r=r_h}\times
     {1\over 2} \exp{\left [{1\over 2}
     \int_{\infty}^{r}<B| T_r^r-T_t^t|B>{\tilde r
     \over b(\tilde r)} d\tilde r \right ]}\; ,
     \end{equation}
which diverges at $r_{h}$ since the exponential is singular at $r_{h}$.
Explicitly, we find that $\kappa$ diverges as $x^{-1/3}$ as $x$ tends to
zero. In addition, a naive application of our
results indicates that the metric is degenerate at the the stretched
horizon since
 \begin{eqnarray}
\sqrt{-g}\, (r_{h}+x)
&=&r^2\sin\theta \exp\left\{\int_{\infty}^{{r_{h}+x}}
{<B|T_r^r-T_{t}^{t}|B>^{ren}
{\tilde r\over 2b(\tilde r)}d\tilde r}\right\}\nonumber \\
&\propto&\left({x\over M}\right)
^{-(1/3+\sqrt{3\alpha}\ln(3\alpha/64)/24+\dots)}\; .
\end{eqnarray}
Both the degeneracy of the corrected metric and the infinite surface
gravity at the stretched horizon are a result of our perturbative
calculation failing in that region and tell us nothing about the
true situation near the unperturbed event horizon.

The breakdown in our calculation might be taken as support for
the viewpoint\cite{s} that
quantum gravity effects need to be included to obtain sensible results
near the event hozion. A more conservative conclusion would be that the
complete, time-dependent back-reaction calculation should be performed.
The time-dependent approach could properly include Hawking radiation
as well as have a sensible perturbative treatment. For instance, in the
Unruh vacuum, $<T_{\mu\nu}>$ is finite since the divergence in the static
vacuum polarization is precisely canceled by the contribution from the
outgoing flux of Hawking radiation~\cite{can}.

What we can say for certain is that we found no evidence that vacuum
polarization gives rise to a short-distance cut-off.

\section*{Acknowledgements}
NJC is grateful for the support provided by a Canadian Commonwealth
Scholarship. JJL appreciates the additional support of the
Jeffrey L. Bishop Fellowship. We thank John Daicic for checking our
expression in Eqn.(\ref{5}).

\end{document}